\documentstyle[twocolumn,aps,epsf,floats]{revtex}
\begin{document}
\draft

\title{Equilibrium and Driven Vortex Phases in the Anomalous Peak Effect}

\author
{J. E. Berger, S. J. Smullin, W. L. Karlin, and X. S. Ling$^\ast$}
\address{
Department of Physics, Brown
University, Providence, Rhode Island 02912}
\medskip
\author{D. E. Prober}
\address
{Departments of Applied Physics \& Physics, Yale University, New
Haven, Connecticut 06520-8284
\medskip\\
\date{\today}\medskip
\parbox{14cm}{\rm
We report a crucial experimental test of the present models of the peak effect in weakly 
disordered type-II superconductors.  Our results favor the scenario in which the peak effect 
arises from a crossover between the Larkin pinning length and a rapidly falling elastic length 
in a vortex phase populated with thermally excited topological defects.  A thickness dependence 
study of the onset of the peak effect at varying driving currents suggests that both screw and 
edge dislocations are involved in the vortex lattice disordering. The driven dynamics in 3D 
samples are drastically different from those in 2D samples. We suggest that this may be a  
consequence of the absence of a Peierls potential for screw dislocations in a vortex line 
lattice.
}}
\maketitle

\narrowtext

In type-II superconductors, vortex lines are embedded in an atomic solid, and thus
the low-temperature condensed 
vortex phase cannot exhibit true long-range
translational order due to the random impurities in the atomic lattice \cite{la}.
 Nevertheless when the random pinning is sufficiently weak, vortex lattice order
can extend to very large distances. Upon increasing temperature or field, the
vortex lattice will eventually transform into a disordered liquid [2-6].  In the
search for the signature of a vortex solid-liquid transition in high-$T_c$
superconducting YBa$_2$Cu$_3$O$_{7-{\delta}}$ (YBCO) crystals, with
increasing temperature and in a strong magnetic field, two
striking phenomena were discovered: a sharp peak in the critical current, {\it
i.e.} a peak effect \cite{lb,da,kwok,lbv}, suggesting a sudden enhancement of vortex pinning;
and, just above the peak effect, a sharp rise \cite{gb} of the ohmic resistance
indicating a sudden increase in the vortex mobility. How
these two effects relate to the microscopic processes by which a (quasi- \cite{bragg}) ordered vortex
lattice transforms into a disordered vortex liquid is still an unsettled issue.

The conventional interpretations of the peak effect are, in one form or the
other, based on Pippard's proposal of a {\it static disordering} of the
vortex lattice in a random pinning potential\cite{pi,lo,ker}.  The peak effect is a crossover
from a regime where the vortex lattice is stiff to a regime where the vortex
lattice becomes too soft to resist the random distortions due to the random pins.
The suggested mechanisms \cite{pi,lo} for softening are of a purely {\it
electromagnetic} origin. Since the discovery of the peak effect in high-$T_c$
superconducting YBCO crystals, the question arises as to whether there is a {\it
thermodynamic} mechanism for the peak effect, and whether the peak effect is
associated with the vortex solid-liquid transformation in the weak random pinning
potential.  For example, the peak effect in YBCO has been described as a result
of {\cite{lmv,tlbc} or a precursor to \cite{kwok} the melting
of the vortex lattice. Specifically, it was argued by Tang {\it et al.} that
within the Larkin-Ovchinnikov theory \cite{lo} of collective pinning, in which 
the {\it typical} pinning force density
is determined by
the shortest elastic length scale in the vortex lattice,
a vortex-lattice melting can give rise to a pronounced peak effect, {\it if} the
melting is more gradual than a one-step process \cite{tlbc}. The rise of the
critical current at the onset of the peak effect is a crossover between two
length scales: the Larkin pinning length and a short-range correlation length in
the intermediate vortex phase(s) determined by {\it thermal} fluctuations, {\it e.g.}
unbound dislocation loops \cite{mn}, or other types of topological
defects \cite{frey}. In this model, the peak effect and the subsequent resistive
transition are different stages of the transition from a (quasi- \cite{bragg}) 
ordered vortex-line
lattice to a completely disordered vortex liquid.
These intriguing possibilities give a strong
motivation for the recent studies \cite{bh,rw,ha,bi,lbp} of the peak effect in
conventional type-II superconductors.

In this Letter, we describe an experiment which, for the first time, tests the
predictions of the classical Pippard-type model and that of Tang {\it et al.} for
the peak effect. For a given weak-pinning sample exhibiting a pronounced peak
effect, the Tang model predicts that after the pinning has been artificially enhanced,
the onset of the peak effect should move {\it upward} in field since the
Larkin length is now reduced and the crossover should occur at a higher field deep
within the intermediate vortex phase(s). The Pippard-type model gives
exactly the {\it opposite} prediction: the onset of the peak effect should
occur at a lower field. Our results are
strongly in favor of the Tang model over the classical Pippard scenario of a
static disordering for the peak effect.  A detailed analysis of the data suggests
that at least two types of topological defects, {\it i.e.} screw and edge dislocations,
are responsible for disordering the vortex lattice and leading to a peak effect.

The experiments are performed on a 2H-NbSe$_2$ crystal ({\char 92}XV1-2" in
\cite{lbp}, renamed as No.2) which was shown previously to exhibit a very
pronounced peak effect, and which has a very low critical current density, demonstrating
the weak-pinning nature of the starting sample \cite{lbp}. The details of the
experimental setup and calibrations are described previously \cite{lbp}.  Except
for the data in Fig.3, all measurements are performed with the samples immersed in a
liquid helium bath with temperature regulated at 4.20 K.

\begin{figure}
\epsfxsize=7.0cm
\epsfbox{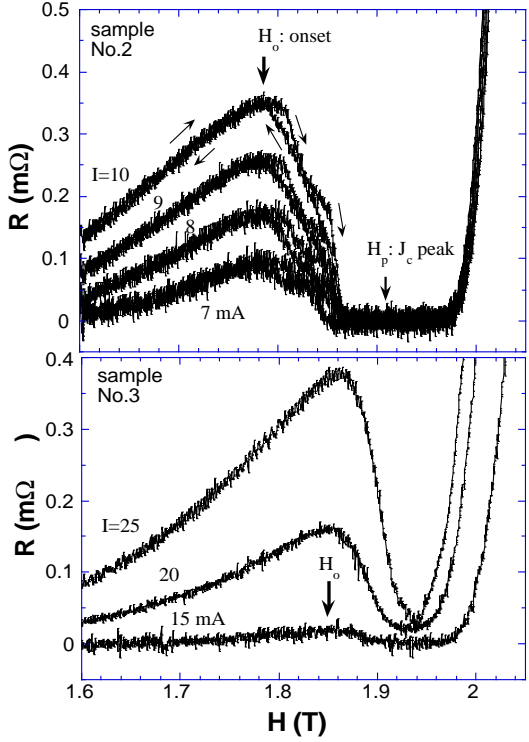}
\vspace{.4cm}
\caption{Top panel: Resistance vs. field at low driving currents 7, 8, 9, and 10
mA for the virgin sample No.2. The thin arrows indicate the direction of the
field sweeps. $H_0$ is the onset of the peak effect; $H_p$ is the apex of the
peak effect ($J_c$ peak).  The region between $H_0$ and $H_p$ is defined as the
{\it peak-effect regime}.  Lower panel: Resistance vs. field at low driving
currents 15, 20, and 25 mA for the cold-worked sample No.3.  The thick arrow
indicates the new onset field of the peak effect.}
\label{f1}
\end{figure}

Fig.1 shows the main result of this paper.   In the top panel, we plot the sample
resistance as a function of the applied magnetic field of the virgin sample
(No.2, dimensions 1.38mm (l) x 1.14mm (w) x 0.015mm (t) \cite{foot1}) at
low driving currents.  The onset of the peak effect is defined as the point at
which the sample resistance starts to {\it decrease} with increasing field.  If
one uses a voltage criterion to determine a critical current density $J_c$, $J_c$
starts to {\it rise} with increasing field at about the same field value (see
Fig.2, open circles). This sample (No.2) exhibits unusual features in the
peak-effect regime.  From $H_0$ to $H_p$, the resistance is hysteretic upon field
cycling. The resistance is higher for increasing field than for decreasing field.
The difference in resistance between the two field cycles displays two striking
peaks \cite{lbp}, suggesting (at least) two kinds of processes in the breakup of
the vortex lattice. This is directly observable from Fig.1(top) in the peak-effect 
regime.

\begin{figure}
\epsfxsize=7.0cm
\epsfbox{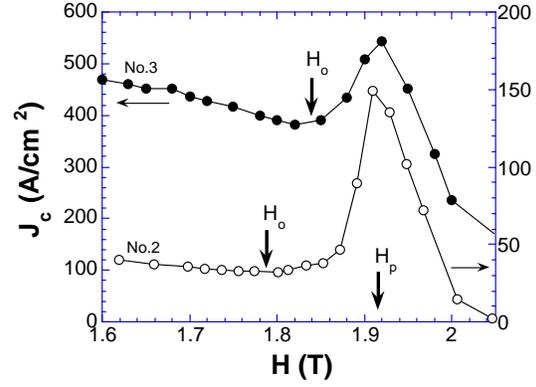}
\vspace{.4cm}
\caption{Critical current density vs. field for samples No.2 and No.3.  The $J_c$
is determined from the $I-V$ characteristics at a voltage criterion of 1 $\mu$V.}
\label{f2}
\end{figure}

We find that, by mechanically bending or cutting the sample, one can add pinning
centers to the sample and increase the critical current density, as shown in
Fig.2. Sample No.3 (dimensions 1.38 x 0.25 x 0.015mm) was cut from the virgin crystal
(No.2).  

From Fig.2, one notices that the $J_c$ at 1.800 T is enhanced by a factor of
12.2, while the $J_c$ at 1.915 T (at the $J_c$ peak) is only increased by a
factor of 3.5, from sample No.2 to sample No.3 after cold-working. This can be
understood in the Larkin-Ovchinnikov theory if the vortex lines are pinned {\it
collectively} below the onset of the peak effect, but {\it individually} at the
apex of the peak effect. In the collective pinning regime, the Larkin-Ovchinnikov
theory predicts that the pinning force density $J_cB \sim
(nf^2)^2 / C_{66}^2 C_{44} r_p^3 \sim n^2$, where $n$, $f$, $r_p$ are
density, force, and size of the pins, respectively. In the individual
pinning regime, at the $J_c$ peak, $J_cB \sim nf \sim n$.  The $J_c$
enhancement factors can be understood if the density of pins in sample No.3 is
3.5 times greater than that of No.2.  A factor of $3.5^2 \sim 12$ increase in $J_c$
is, therefore, expected in the low-field collective pinning regime. This analysis also
explains why the peak effect is more pronounced in clean samples, since the ratio
peak-$J_c$/onset-$J_c \sim n^{-1}$ increases with decreasing pin density. A
transmission electron microscopy examination \cite{paine} of a similar
cold-worked sample reveals large numbers of dislocations in the atomic lattice of
the 2H-NbSe$_2$ crystal, suggesting that these mechanically induced
atomic-lattice defects are responsible for the enhanced pinning.

Thus the peak effect data in Fig.2 can be explained by the Larkin-Ovchinnikov
theory if one assumes a crossover from a collective pinning regime below the peak
effect to an individual pinning regime within the peak effect region. The underlying
mechanism of this crossover, however, is the issue of current interest.

Larkin and Ovchinnikov suggested a mechanism \cite{lo} whereby this crossover
occurs when the transverse Larkin length, $R_c$, becomes smaller than the effective
penetration depth $\lambda^{'} = \lambda_L/(1-b)^{1/2}$, where
$\lambda_L$ is the London penetration depth, and $b=B/H_{c2}$.  When this
occurs, the nonlocality of $C_{44}$ becomes important and provides additional vortex-lattice
softening.  Thus R$_c$ drops further and $C_{44}$ becomes softer
still, leading to an elasticity ``ultraviolet catastrophe.'' Since $R_c \sim
C_{66}^{3/2}C_{44}^{1/2}r_p^2/nf^2$, adding
disorder ($nf^2$) reduces $R_c$, and the crossover should occur at a lower field.
 This is exactly the {\it opposite} of what we observe here. If Fig.2 is somewhat
less convincing, Fig.1(lower panel) shows clearly that the onset of the peak
effect in the cold-worked sample No.3 is higher than that of virgin sample No.2. 
In addition, the fine structures of the peak effect in the virgin sample are no
longer observable. As shown in Fig.1(lower panel), in sample No.3, the resistance
drop is essentially featureless and is no longer hysteretic in the peak-effect
regime.

The disappearance of the fine features in the peak-effect regime and the upward
shift in the onset field of the peak effect, however, are what one would expect
if the onset of the peak effect in the virgin sample is an elastic instability
driven by {\it thermal} fluctuations. In the Tang model \cite{tlbc}, the onset of
the peak effect is a crossover from the Larkin pinning length $R_c$ to a rapidly
decreasing (with increasing $T$ or $H$) elastic length set by thermal fluctuations
(although the random pinning may still play a role in the the elastic instability
and the new elastic length). In the peak-effect regime, the vortex lattice behaves
like a liquid at large length scales, but behaves like a solid at a short length
scale.  The long-wavelength liquid-like property of the pinned vortex lattice is
cut off by the Larkin pinning length, and cannot manifest in the vortex transport
until the liquid-like correlation length becomes smaller than the
Larkin pinning length \cite{tlbc}. In this picture, by adding disorder ($nf^2$) 
to a clean sample, the
Larkin length, $R_c$, becomes smaller, and the crossover occurs at a higher
field, as is seen here.  We thus propose that there is an equilibrium phase
transition in the vortex lattice at the onset of the peak effect in the clean
sample (No.2) where the resistance hysteresis begins. The pinned vortex lattice
starts to break up spontaneously at this transition, but it takes two stages
(each of which seems to be a first-order transition) for the vortex lattice to
become completely disordered at the apex of the peak effect, both of which occur
in a very narrow region of the $H-T$ diagram (see Fig.3 below).  An {\it apparently}
flowing vortex line fluid, characterized by an {\it observable} ohmic resistance, 
is achieved only at a larger (disorder-dependent) field, 2.05 T for sample No.2 and
2.40 T for sample No.5.

Here we should point out an early study of the effects of added pins on the peak
effect in V$_3$Si crystals \cite{kupfer}.  Although it was not discussed, the
$J_c$ vs. $H$ data showed that the onset of the peak effect shifted {\it upward} in field at
low irradiation dosages.  Further increases in irradiation shifted the onset to
lower fields, leading to broad $J_c$ peaks as a function of magnetic
field. Similar broad $J_c$ peaks were also found in high-$T_c$ superconductors,
known as the ``fish-tail'' effect \cite{larb} or the ``second-peak'' effect
\cite{zel}.  However, if one plots the onset fields of these latter effects on
the $H-T$ phase diagram, they are either temperature independent or have a positive
slope \cite{zel}.  Thus $J_c$ in those samples is a monotonic function of temperature.
 In Fig.3, 
we plot the
temperature dependence of the onset of the peak effect on the $H-T$ phase diagram
for sample No.3 (obtained using a variable temperature cryostat).  A
similar phase diagram was also measured using an ac-susceptibility technique
\cite{grover}. Because its phase line has a negative slope on the $H-T$ diagram,
the peak effect here can be observed as a function of temperature and magnetic
field.  Since the ``fish-tail'' \cite{larb} or the ``second-peak'' \cite{zel} effects
 are consistent with the $C_{44}$ nonlocality
catastrophe scenario ({\it e.g.} the onset shifts downward in field with increasing
disorder \cite{kupfer}) and are observable even with strong pinning \cite{kupfer}, 
one may dub this behavior a
{\it normal peak effect}.  The peak effect in weak-pinning samples,
observed as a function of both $T$ and $H$, may be termed an {\it anomalous peak 
effect}.  The two effects appear to have different mechanisms, 
as also pointed out by Larkin {\it et al.} \cite{lmv}.
This distinction may help settle many unresolved issues in both phenomena.
 
\begin{figure}
\epsfxsize=7.0cm
\epsfbox{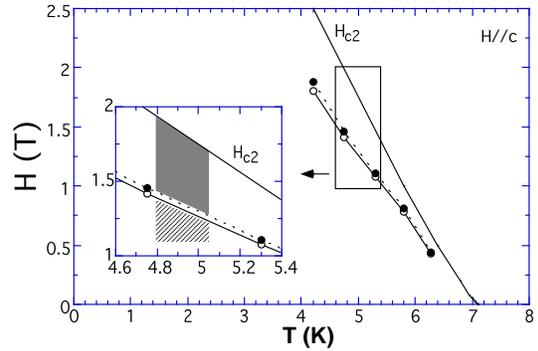}
\vspace{.4cm}
\caption{Phase diagram of an {\it anomalous peak effect} (see text for
definition): open circles are the onsets of the peak effect; filled circles are
the resistance dips ($H \parallel$ c, sample No.3).  The upper critical field $H_{c2}(T)$
line (determined by magnetization) is taken from the literature [29].  The
line-shaded region indicates a pinned vortex lattice, the upper shaded region
indicates a disordered vortex phase.}
\label{f3}
\end{figure}

Since the onset of the peak effect is where the short-range ordered vortex lattice
becomes even more disordered, one can use the current dependence of this onset
to study the driven vortex phases and the dynamic phase transitions \cite{bh,kv,hel} between them.  
It turns out that, as shown below, the driven dynamics of the vortex lattice can also shed new light
into the origin of the peak effect and the nature of vortex-lattice disordering ({\it e.g.}
screw or edge dislocations, etc.).

\begin{figure}
\epsfxsize=7.0cm
\epsfbox{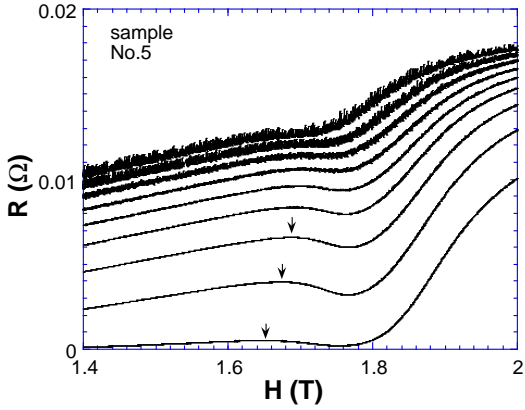}
\vspace{.4cm}
\caption{Resistance vs. magnetic field at driving currents: 90 to 10 mA (from top
to bottom, 10 mA apart). $T = 4.20$ K, $H \parallel$ c.  The arrows indicate the onset H$_0$
of the peak effect.  Sample No.5: thickness d = 0.75$\mu$m.}
\label{f4}
\end{figure}

Fig.4 shows the resistance vs. field for different driving currents for 
sample No.5 (0.75$\mu$m in thickness). (The noise at high driving
current may be caused by the shunt current in the silver paint voltage contacts
since the sample is very thin.)  The sample No.5
was peeled off from sample No.2 using 
clear sticky tape (the tape can be dissolved
in toluene) and has dimensions $\sim$1.0mm x 0.5mm x 0.75$\mu$m.  
For this thin sample, a well defined, although
broad, resistance dip and, correspondingly, a shallow peak in $J_c$, can be
observed.  The ratio peak-$J_c$/onset-$J_c$ is $\sim$ 1.1, and the peak $J_c$ (2200 A/cm$^2$ at 1 $\mu$V)
 is at 1.73 T. The measured phase line (not shown) of the peak effect in sample 
 No.5 still has a negative slope, similar to Fig.3.
   An order-of-magnitude estimate \cite{kor}
suggests that the longitudinal Larkin length $L_c$ is much greater than the
thickness of the sample across the peak-effect regime, up to 1.80 T in this sample.  We thus believe that the 
higher $J_c$ and the lower field of the peak effect in sample No.5 are results of the 2D nature
of the sample.

Again, if we define the onset of the peak effect at the resistance
maxima (when still discernable), a dynamic phase diagram of the peak effect can be
constructed, as shown in Fig.5. There are two major differences between the
dynamic phase diagrams in thick and thin samples. For low driving currents, the
onset of the peak effect in thick samples seems to settle at a finite large
magnetic field value,  $\sim$ 1.78 T for sample No.2 (Fig.5 inset, a replot of Fig.4b in \cite{lbp}) and 
$\sim$ 1.85 T for sample No.3 (see Fig.1).  However, for the thin sample No.5,
the onset of the peak effect appears to continuously decrease in field with
decreasing current. In fact, the data can be fitted to an empirical form $I \sim
1/(H^{\ast}-H_0)^2$, implying that there is no end point on the field axis for the
onset of the peak effect in the limit of small driving current.  For large and
increasing driving currents, the onset of the peak effect in thick samples tends
to turn around and decrease rapidly.  In thin samples, it continuously
increases in field until the peak effect (resistance dip) is no longer
identifiable.  We suggest that this difference in driven
dynamics arises from the dimensionality of the system.

\begin{figure}
\epsfxsize=7.0cm
\epsfbox{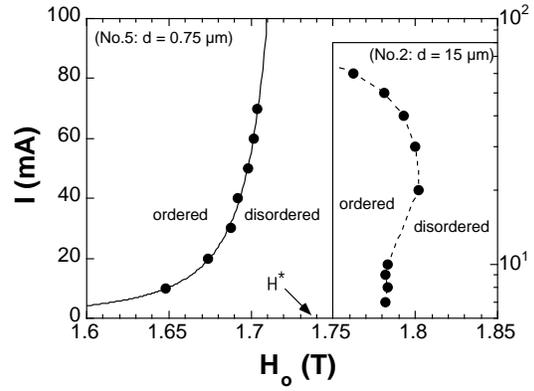}
\vspace{.4cm}
\caption{Dynamic phase diagram in 2D: driving current vs. onset field H$_0$ of
the peak effect for sample No.5, $H \parallel$ c, $T = 4.20$ K. The solid line is a fit to $I
\sim 1/(H^{\ast}-H_0)^2$, where $H{^\ast} = 1.737$ T.  Inset: Dynamic phase diagram in
3D: sample No.2 [23].  The dashed line is a guide to the eyes. Note that the current
axis is on a log-scale.}
\label{f5}
\end{figure}

Since 
edge dislocations are the most likely topological defects in a 2D lattice, we interpret
the observed peak effect in 2D samples as a result of an elastic length crossover 
from the Larkin length below
the peak effect to a rapidly decreasing elastic length set by thermally excited
edge dislocations \cite{tlbc}.  The lack of a lower end point on the $I$ vs. $H_0$ curve
(Fig.5) is consistent with the common belief \cite{foot2} and numerical
simulations \cite{gh} that edge dislocation pairs are always unbound at arbitrary
temperature in a random potential, in 2D.

If the low current part of the dynamic phase diagram suggests an unstable 2D
vortex lattice at rest, the higher current behavior would suggest an unstable 3D
vortex lattice in motion. Apart from the testable, trivial explanations (such as
self-heating, suppression of superconduting order parameter, etc.), which have
been ruled out, this result may suggest that the different dynamics of screw and
edge dislocations both play a role.  In a 3D vortex-line lattice, there can be both screw and
edge dislocations, or combinations ({\it e.g.} loops) of the two.  Unlike the edge
dislocations in a 2D lattice, which are always immobilized by a Peierls
potential, the screw dislocations are much more mobile since there is no discrete
periodicity along the field direction \cite{br2}. Thus, when pairs of screw
dislocations are generated, due to the interaction between the moving lattice and
the pins, there are two competing processes.  They recombine, if the motion is
slow and the rate at which they are generated is low.  However, if the rate of
generation is higher than the rate of recombination, a Kosterlitz-Thouless-$like$
unbinding transition \cite{kos} may occur and result in a moving disordered vortex lattice
full of free screw dislocations.

We are grateful to A. Houghton, D. A. Huse, M.C. Marchetti, T. Nattermann, S. Ryu, and
Chao Tang for helpful discussions, and to J.M. Valles for a critical reading of
the manuscript and for forcing us to clarify several important points. S.J.S. is an undergraduate UTRA Fellow.
\smallskip

\smallskip

$^\ast$An Alfred P. Sloan Research Fellow and the corresponding author. 
Electronic address: xsling@brown.edu.


\begin{references}
\bibitem{foot1}
The thickness of the sample No.2 was inaccurately quoted as 20 $\mu$m in
\cite{lbp}. It should be 15 $\mu$m according to a SEM measurement.
\end{references}
\end{document}